\documentclass[letterpaper, 10 pt, conference]{ieeeconf}  

\IEEEoverridecommandlockouts                              
\usepackage{graphicx} 
\usepackage{amsmath} 
\usepackage{amssymb}  
\usepackage{xcolor}
\usepackage{lipsum}
\usepackage{multirow}
\usepackage{cite}
\usepackage{hyperref}
\hypersetup{
    colorlinks=true,
    linkcolor=black,      
    urlcolor=blue,
    }
\title{\LARGE \bf
Wireless Earphone-based Real-Time Monitoring of Breathing Exercises: A Deep Learning Approach   
}

\author{Hassam Khan Wazir, Zaid Waghoo, and Vikram Kapila
\thanks{The authors are with the Mechanical and Aerospace Engineering Department, NYU Tandon School of Engineering, Brooklyn, NY 11201, USA (corresponding author: 646-997-3161; e-mail: vkapila@nyu.edu).}
}

\begin{document}

\begin{figure*}[!t]
\vspace*{-18cm}
\noindent © 2024 IEEE. Personal use of this material is permitted. Permission from IEEE must be obtained for all other uses, in any current or future media, including reprinting/republishing this material for advertising or promotional purposes, creating new collective works, for resale or redistribution to servers or lists, or reuse of any copyrighted component of this work in other works.
\end{figure*}


\newpage

\maketitle
\thispagestyle{empty}
\pagestyle{empty}

\begin{abstract}
Several therapy routines require deep breathing exercises as a key component and patients undergoing such therapies must perform these exercises regularly.
Assessing the outcome of a therapy and tailoring its course necessitates monitoring a patient's compliance with the therapy. 
While therapy compliance monitoring is routine in a clinical environment, it is challenging to do in an at-home setting.
This is so because a home setting lacks access to specialized equipment and skilled professionals needed to effectively monitor the performance of a therapy routine by a patient.  
For some types of therapies, these challenges can be addressed with the use of consumer-grade hardware, such as earphones and smartphones, as practical solutions.
To accurately monitor breathing exercises using wireless earphones, this paper proposes a framework that has the potential for assessing a patient's compliance with an at-home therapy.
The proposed system performs real-time detection of breathing phases and channels with high accuracy by processing a $\mathbf{500}$ ms audio signal through two convolutional neural networks.
The first network, called a channel classifier, distinguishes between nasal and oral breathing, and a pause.
The second network, called a phase classifier, determines whether the audio segment is from inhalation or exhalation.
According to $k$-fold cross-validation, the channel and phase classifiers achieved a maximum F1 score of $\mathbf{97.99\%}$ and $\mathbf{89.46\%}$, respectively.
The results demonstrate the potential of using commodity earphones for real-time breathing channel and phase detection for breathing therapy compliance monitoring.

\vspace{0.5em} 
\indent \textit{Clinical relevance}—This paper introduces a real-time monitoring system for breathing that can facilitate therapy compliance for several breathing-based exercises.

\vspace{0.5em} 

\end{abstract}

\section{INTRODUCTION}

Deep breathing exercises have been shown to reduce conditions such as hypertension \cite{herawati2023} and anxiety \cite{goessl2017}.
Moreover, several therapy interventions for medical conditions such as lymphedema \cite{fu2014}, asthma \cite{thomas2008, bruton2018}, and chronic obstructive pulmonary disease \cite{lu2020} rely on breathing exercises as a primary component.
Thus, tracking breathing using a monitoring device can support therapy compliance. 

Respiratory rate, blood carbon dioxide level, and the volume of air inhaled and exhaled by the lungs serve as vital indicators of health. 
In a clinical setting, these indicators are measured using devices such as spirometers and capnographs \cite{liu2019}. 
However, for general diagnosis and therapy compliance applications, metrics such as the respiratory rate are often measured manually by counting the number of chest expansions over a minute \cite{liu2019}.
For such non-acute applications, several devices are commercially available including respiratory thoracic belts and pulse oximeters with integrated respiratory rate monitoring \cite{liu2019}.
A major drawback of these devices is that they are considered specialized medical hardware that may not be readily available at home.
To address these issues, one viable alternative is the use of commodity hardware to measure acoustic biomarkers for monitoring therapy compliance.

Prior research has explored the effectiveness of acoustic data for administering therapies such as breathing training \cite{shih2019} and for tracking metrics such as respiratory rate \cite{doheny2023, romano2023}.
These studies rely on a user breathing directly into a microphone, which limits their application for at-home therapy compliance monitoring, especially when the therapy requires physical movement with breathing exercises.
These studies also do not distinguish between nasal and oral breathing, which is important for some therapy interventions \cite{fu2014}.

Recent advances in acoustic hardware design have led to the development and, consequently, the widespread adoption of wireless earphones that are capable of connecting to smart mobile devices.
Almost all wireless earphones have a built-in microphone and several latest models
also include inertial measurement units (IMUs) and proximity sensors.
These developments open new sensing modalities and research avenues for digital healthcare and telerehabilitation.
For example, recent research has used wireless earphones for respiratory rate monitoring using audio \cite{kumar2021} and the built-in IMU \cite{roddiger2019} sensors of the earphone.
Yet other studies have combined prior approaches to integrate the audio signals and IMU data from the earphones for determining the respiration rate and breathing channel (nasal \textit{vs.} oral breathing) \cite{rahman2022,islam2021}.
All of these studies utilize deep-learning models to detect respiratory rate and  breathing channel.
It is pertinent to note that the aforementioned studies do not provide the dataset used to train their underlying neural network models, which makes it impossible to reproduce their results.

To accurately monitor breathing exercises using wireless earphones, this paper creates a framework that has 
the potential for assessing a patient's compliance with an at-home therapy.
The study builds upon \cite{wazir2020} that developed an at-home lymphedema therapy compliance monitoring system.
Along with physical activity, such therapies entail breathing exercises wherein a user performs nasal inhales and oral exhales, which warrant accurate tracking of breathing phases (inhale/exhale) and channels (nasal/oral).
Two main contributions of this study are: (i) a system for real-time detection of breathing phases and channels when the user performs breathing exercises and (ii) an annotated breathing audio dataset, created using wireless earphones, to facilitate the reproducibility of results and enable further research.

The paper is organized as follows. Section \ref{sec:designanddevelopment} elaborates on the creation of the dataset and the architecture and training of the system used for breath classification. Section \ref{sec:resultsanddiscussion} discusses the results based on the preliminary work and addresses the benefits and shortcomings of the approach used. Finally, Section \ref{sec:conclusion} provides concluding remarks and suggests directions for future research.

\section{DESIGN AND DEVELOPMENT} 
\label{sec:designanddevelopment}

The proposed system is envisioned as an application (App) running on a smart mobile device.
Using a pair of wireless earphones connected to the smart device, the system will receive audio signals in real-time and use them to infer the breathing pattern of the user. 
Specifically, the App will detect (i) the breathing channel, i.e., nasal or oral, and (ii) the breathing phase, i.e., inhale or exhale. 
To begin using the system, the user will launch the App on the smart device and wear the wireless earphones. 
Next, 
the App will acquire the breathing audio recorded through the wireless earphones and pass it through a neural network classifier that will infer the channel and phase of breathing. 
Additionally, the classifier will detect any pause between breathing and, thus, can be used to calculate metrics such as the total number of breaths, respiratory rate, and breath-phase duration.
Below we provide details about a proof-of-concept system where the data-gathering process and the architecture of the neural network have been implemented on an Apple iPhone and a high-performance computing system, respectively. Porting the designed neural network to the iOS environment will enable the App-based system.

\subsection{Audio dataset creation} 
\label{sec:audiodatasetcreation}

As of this writing, datasets of breathing audio sounds recorded with wireless earphones are not publicly available.
Moreover, a majority of the publicly available breathing audio datasets are either used for detecting pathologies of the respiratory tract \cite{bhattacharya2023} or were recorded using contact microphones \cite{rocha2018}. 
Finally, the data quality of many public datasets is too low due to signal corruption from noise, distortions, and audio artifacts, making the datasets unsuitable for our application.
In response, we sought to create a breathing audio dataset using wireless earphones.

To capture the audio data, a pair of wireless AirPods connected to an iPhone were chosen due to their broad availability and popularity.
Airpods operate under three modes, i.e., noise cancellation, transparency, and off.
To eliminate any unwanted audio interference or amplification caused by the noise cancellation and transparency modes, the off mode was used for audio recording. 
The recordings were stored in the M4A audio file format and converted to the WAV format.

For the dataset collection, eight healthy individuals, aged between $23$ and $34$ years, were invited for two recording sessions.
During each session, they were asked to record two-minute-long audio clips each for nasal and oral breathing.
The recordings were done in a quiet room and each participant was given a demonstration of the exercises before the recording session.
Participants who owned AirPods were encouraged to use them, rest of the participants were provided with AirPods and new ear tips.

\begin{figure}[!b]
    \centering
    \includegraphics[width=0.41\textwidth]{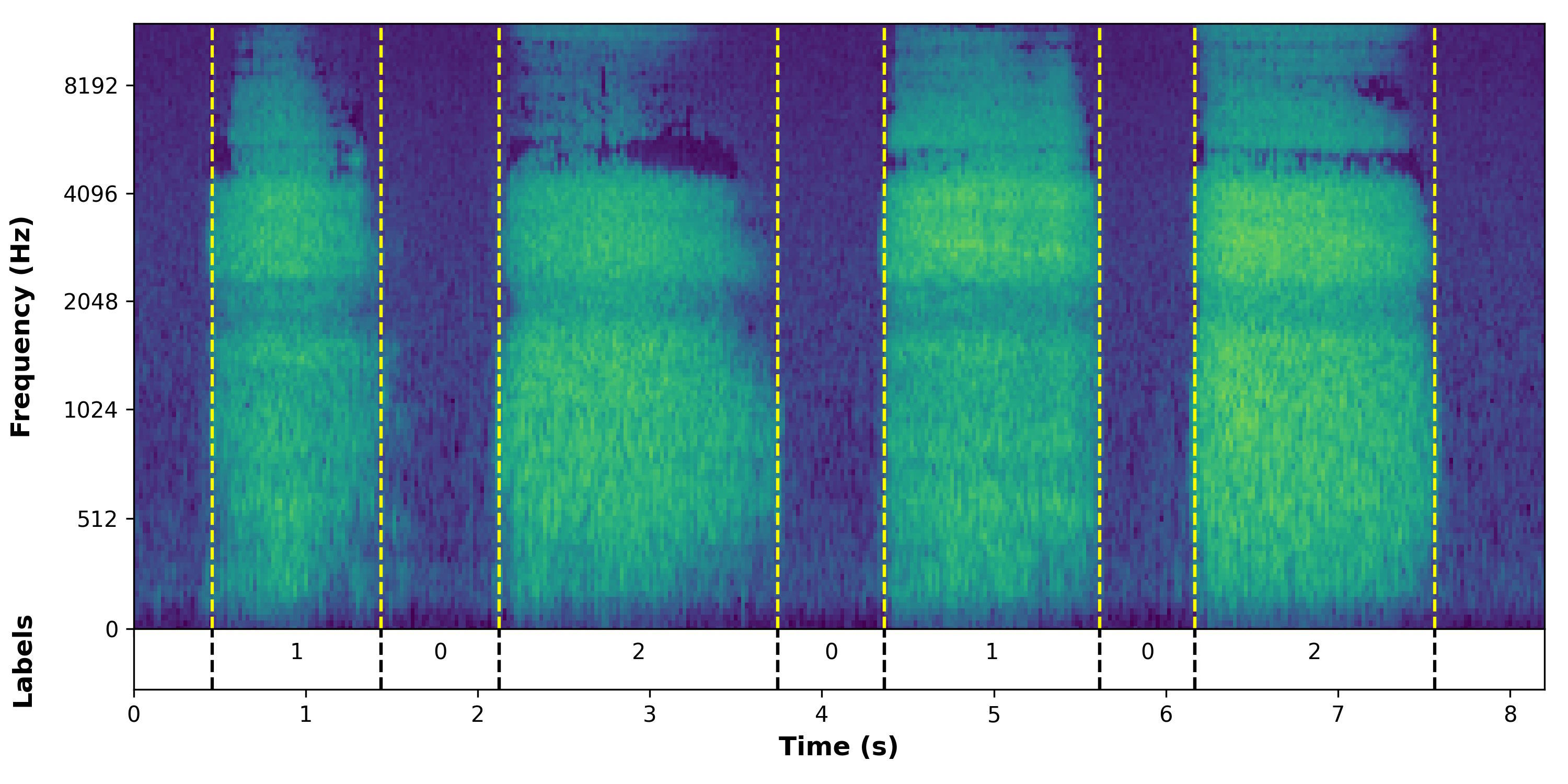}
    \vspace{-0.8em} 
    \includegraphics[width=0.41\textwidth]{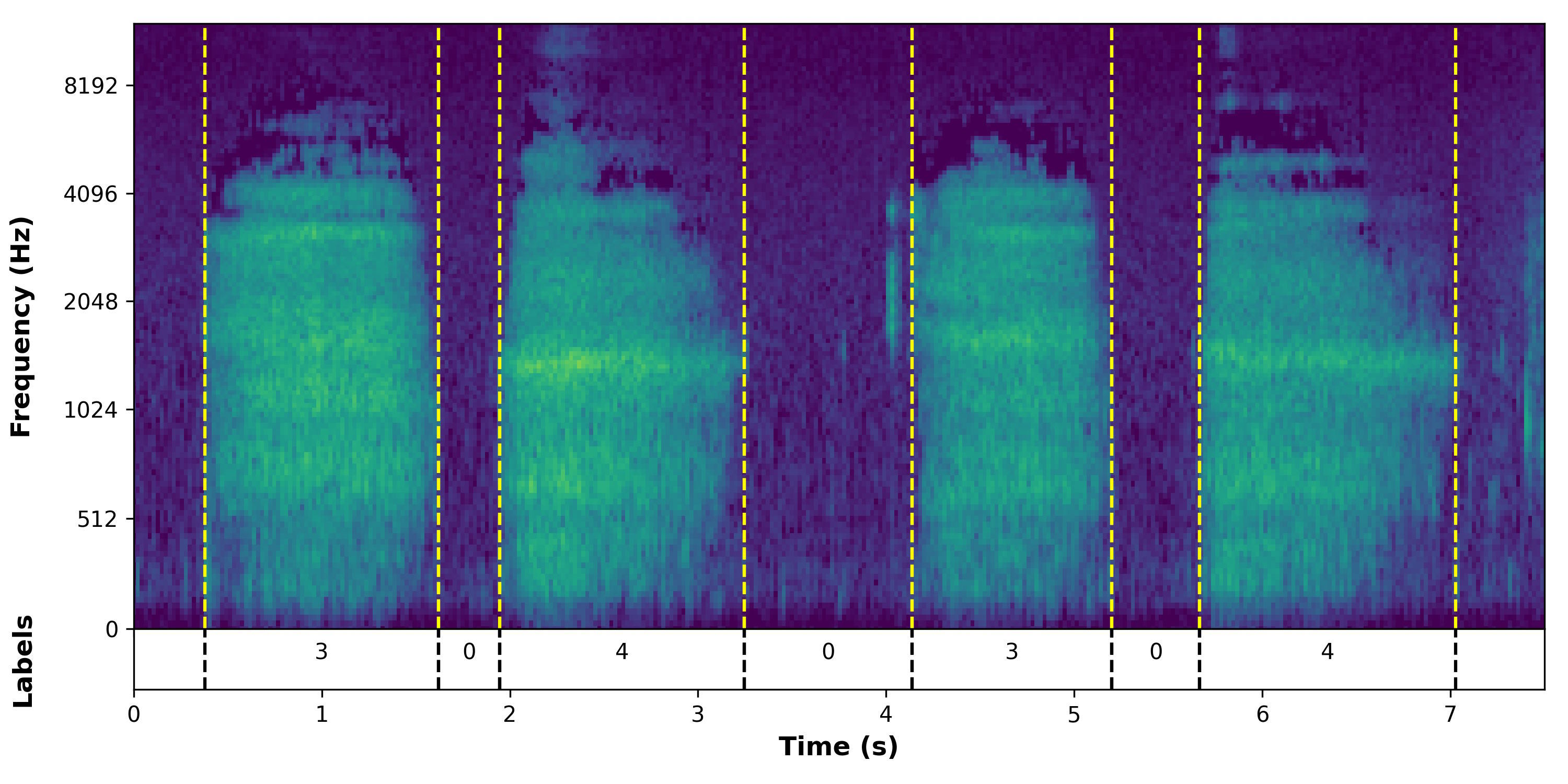}
    \vspace{-0.2em}    
    \caption{Audio spectrograms for (top) nasal and (bottom) oral breathing.}
    \label{fig:labeledExample}
\end{figure}

To label the audio data, an initial dataset was recorded with two participants who performed two sets of two-minute-long nasal and oral breathing exercises.
After being converted to WAV format, the audio files were manually labeled using the open-source application ``Audacity''.
The labeling was performed by visualizing the audio files as spectrograms and listening to the audio using earphones.
The following categories were used for a systematic labeling: pause (0), nose-inhale (1), nose-exhale (2), mouth-inhale (3), and mouth-exhale (4). 
These labels were assigned to the corresponding sections of audio, providing a structured representation of the breathing patterns.
The labels were exported as a text file where each row contains the start time, end time, and label of the audio segment. Figure \ref{fig:labeledExample} shows an example of annotated spectrograms for nasal and oral breathing.

To simplify the labeling for the recordings of remaining six participants, the labeling process was streamlined by training a binary convolutional neural network (CNN) model to distinguish between two classes: pause (0) and breath (${\text{\textasciitilde}}0$).
This model achieved an accuracy of $90\%$, which significantly expedited the labeling process.
The pre-labeled files obtained through CNN underwent a round of scrutiny to rectify any minor mistakes made by the model during inference, and to provide the correct classes ($1, \ldots , 4$),
for the data from the breath class (${\text{\textasciitilde}}0$), 
based on the channel and phase of breathing.
This classifier was re-trained after the addition of each new participant's data until an accuracy of $96\%$ was achieved.
The architecture of this labeling classifier is identical to the channel classifier model discussed in the next subsection with the exception that the labeling classification head was modified to output two classes instead of three given by the channel classifier. 
The dataset is made publicly available at: \url{https://shorturl.at/jlrKU}.

\begin{figure*}[!t!h]
    \centering
    \includegraphics[width=1.0\textwidth]{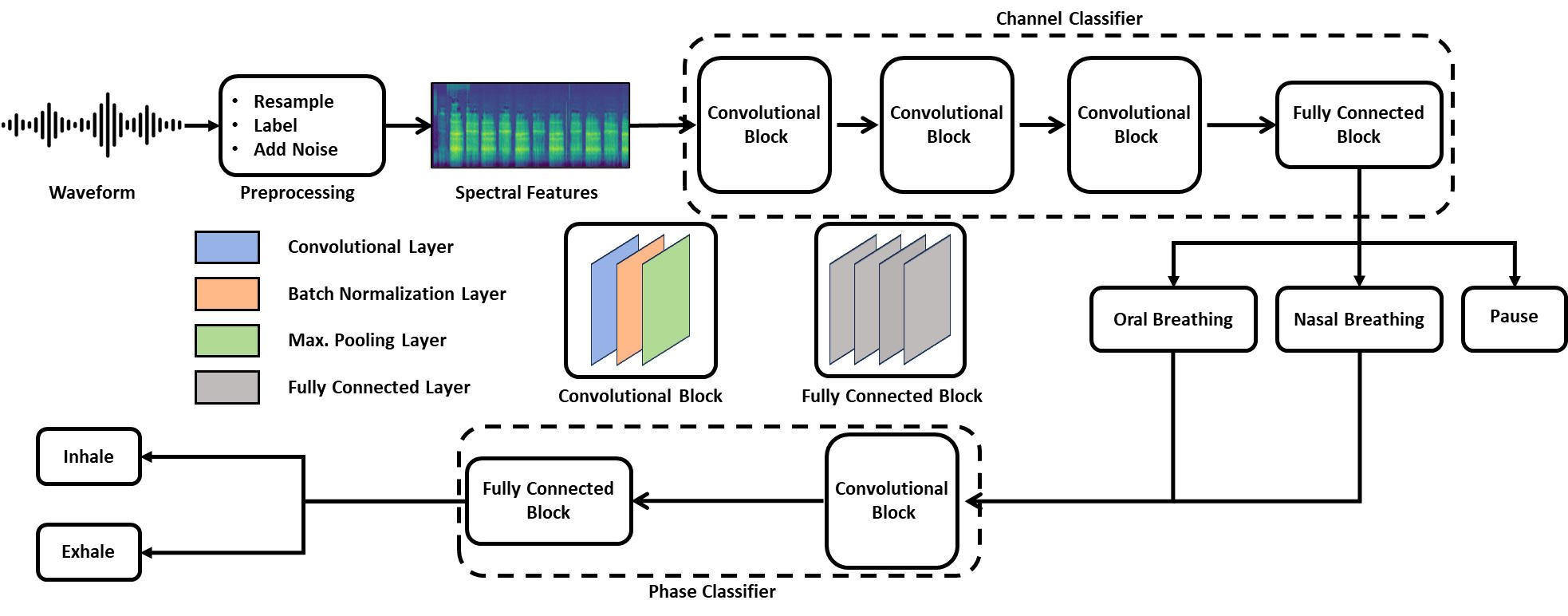}
        \vspace{-2em}
    \caption{The system architecture for detecting pause, breathing channels, and breathing phases.}
    \label{fig:systemarchitecture}
    \vspace{-1.5em}
\end{figure*}

\vspace{0em}
\subsection{Training the model}
\vspace{-0.2em}
The next steps entailed preprocessing the audio data, tuning model hyperparameters, and training the neural network to distinguish between channels and phases of breathing. 

\subsubsection{Audio preprocessing}
The WAV audio clips underwent several preprocessing steps before being used to train the model.
To begin with, the audio clips were resampled at $16$ kHz to standardize the input data.
Next, the audio clips were segmented and labeled into $500$ ms long intervals with a frame stride of $250$ ms to allow overlap between adjacent segments.
The frame length of $500$ ms was deemed to have enough information for model training without adding too much delay to the system during inference.
To label each audio segment, the text file of labels for the corresponding audio clip was used. 
We employed a multilabeling approach wherein the audio segments were allowed to have more than one label.
This is important because an audio segment that contains, e.g., a transition from a pause to a nasal inhale should be assigned labels for pause and nose-inhale.
The labels were also one-hot encoded to prevent the model from learning ordinal relationship between the categories.

To improve the generalizability of the model, the training dataset was augmented with noise taken from an audio clip that includes background noise from multiple people talking.
The noise was uniformly sampled across a range of signal-to-noise ratios from $20$ dB to $40$ dB. 

The segmented audio waveforms were converted to spectral features, i.e., mel-spectrograms and mel-frequency cepstral coefficients (MFCCs) of sizes $128\times126$ and $40\times41$, respectively.
The mel-spectrograms were divided into $128$ mel-filterbanks, with 2048 fast fourier transform points, a window length of $2048$, and a hop length of $64$.
To create the MFCC matrices, the number of MFCCs was chosen to be $40$.
The mel-spectrograms and MFCCs were generated using the TorchAudio \cite{yang2021} library from PyTorch.

\subsubsection{Model training pipeline}

The problem statement for model training was split into two tasks.
The first task entailed classifying between pause, nasal breathing, and oral breathing.
To do this, a CNN classifier was designed with three convolutional blocks having filter sizes of $8$, $16$, and $32$, respectively. 
Each convolutional block's convolutional layer was followed by a batch normalization layer and a max pooling layer.
Moreover,  four fully connected layers were added after the last convolutional block.
The neural network used a rectified linear unit as the activation function and each convolutional layer had a $3\times3$ kernel size.
As noted in subection \ref{sec:audiodatasetcreation}, this model was also used to train the binary classifier for data labeling.
The classifier's architecture is shown in Figure \ref{fig:systemarchitecture} and denoted as `channel classifier'.

Having passed through the channel classifier, the audio data was classified as either pause, nasal breathing, or oral breathing. 
If an audio segment was classified as a pause, the system displayed the result to the user and began processing the next audio segment.
However, if the audio segment was classified as nasal or oral breathing, then the audio segment was passed to a second classifier that sought to determine whether it represented an inhale or exhale.
This classifier had a single convolutional block with a filter size of $4$ and a $3\times3$ kernel size.
The convolutional block was followed by four fully connected layers.
The architecture of the CNN is shown in Fig. \ref{fig:systemarchitecture} as `phase classifier'.

A binary cross-entropy loss function was used to train all models. 
In each instance, the model was trained until the F1 score of test dataset stopped improving for $30$ epochs, after which the highest F1 score was reported and the corresponding model checkpoint was saved. 

\vspace{-0.2em}
\section{RESULTS AND DISCUSSION}
\label{sec:resultsanddiscussion}
\vspace{-0.2em}

To evaluate the system performance, we conducted $k$-fold cross-validation on the dataset for providing a better estimate of the model performance on new data.
Specifically, we used a leave-one-out cross-validation (LOOCV) method where data from seven subjects was used for training and the eighth subject's data was used for validation.
This method prevents any data leakage between the training and testing datasets, and it also aids us in identifying any limitations of the model and detecting any shortcomings of the dataset.

With LOOCV of the channel classifier, we observed that using mel-spectrograms as an input feature produced the best results with an average F1 score of $93.98\%\,(\text{SD=}5.02\%)$. 
In comparison, using MFCCs as input features produced an average F1 score of $90.35\%\, (\text{SD=}8.49\%)$.
When using mel-spectrograms, the highest F1 score of $(97.99\%)$ was obtained by excluding $\text{user } \#5$ from the training dataset.
Conversely, the lowest F1 score of $(82.45\%)$ was obtained by excluding $\text{user }\#6$ from the training dataset.
When using MFCCs, the results were similar to the mel-spectrograms case, with the highest F1 score of $(97.50\%)$ (by excluding $\text{user } \#5$) 
and the lowest F1 score of $(75.93\%)$ (by excluding $\text{user } \#2$).

The phase classifier was evaluated using a similar LOOCV approach and a comparison between mel-spectrograms and MFCCs as input features revealed that mel-spectrograms were a better input feature for training the phase classifier.
The average F1 score with mel-spectrograms was seen to be $76.20\%\,(\text{SD=}8.76\%)$ which was higher than the average F1 score of $75.56\%\,(\text{SD=}8.55\%)$ with MFCCs.
The results also had a high standard deviation across the folds.
Using mel-spectrograms, the highest F1 score of $89.46\%$ was obtained by excluding $\text{user } \#4$ from the training dataset, and the lowest score of $63.33\%$ was obtained by excluding $\text{user } \#6$ from the training dataset.
When using MFCCs, the highest and lowest F1 scores obtained were $87.35\%$ (by excluding $\text{user } \#4$) and $63.71\%$ (by excluding $\text{user } \#5$), respectively.

\color{black}
The average F1 scores for phase classification were relatively low in comparison to the average F1 scores for channel classification.
This suggests that either the CNN is not a very effective architecture to classify breathing phases, or there is insufficient variety in the dataset to fully capture the problem's complexity. 
Overall, the mel-spectrograms as input features performed better than MFCCs in training a neural network for breathing channel and phase classification.
Table \ref{tab:results} shows the results of the experiments and provides a comparison between LOOCV using the two input features.

\section{CONCLUSION}
\label{sec:conclusion}

In this paper, we proposed a system for real-time detection of breathing channels and phases by capturing audio signals from a pair of wireless earphones when a user performs breathing exercises.
This work opens future research avenues for the design and development of at-home therapy compliance applications involving breathing exercises.
Our work shows that audio signals can be utilized for applications related to digital health and telerehabilitation using commodity hardware.
Future work will explore the combination of vision-based models with acoustic models to simultaneously detect and track physical and breathing exercises for a holistic compliance monitoring solution.
Such work can lead to the development of a personalized therapy compliance monitoring system that is tailored to specific users, furthering the scope of this work in remote monitoring systems.

\section*{DECLARATION}
\label{sec:declaration}

The data was obtained with the participants' signed consent and the research was approved by the NYU Institutional Review Board (IRB-FY2020-4198).

\begin{table}[!t]
\vspace{1em}    
    \centering
\caption{Comparing mel-spectrograms and MFCCs as input features}
\vspace{-0.5em}
\label{tab:results}
    \begin{tabular}{|l ||c|c | c| c|} 
    \hline
    \multirow{2}{4em}{\textbf{Classifier}}  & \multirow{2}{4em}{\textbf{Features}}  & \multicolumn{3}{|c|}{\textbf{F1 Score $(\%)$}}\\
    \cline{3-5}
                &                           &  \textbf{Avg. (SD)}     &\textbf{Max.}   & \textbf{Min.} \\
    \hline
    \multirow{2}{4em}{\textbf{Channel}}              &   Mel-spectrogram         & $93.98\,(5.02)$             & $97.99$         & $82.45$\\
    \cline{2-5}
                &  MFCC     & $90.35\,(8.49)$      & $97.50$     & $75.93$     \\ 
    \hline
    \multirow{2}{4em}{\textbf{Phase}}    &   Mel-spectrogram         & $76.20\,(8.76)$  & $89.46$             & $63.33$\\
                                                        \cline{2-5}
                &  MFCC     & $75.56\,(8.55)$      & $87.35$     & $63.71$     \\ 
        \hline 
    \end{tabular}
    \vspace{-2em}
\end{table}

\addtolength{\textheight}{-12cm}   

\bibliographystyle{IEEEtrans}
\bibliography{references}{}


\end{document}